\documentclass[a4,12pt]{article}

\usepackage{amsmath,amssymb,amsfonts,graphics,graphicx,amscd,amsfonts,epsf,epsfig,color}
\makeatletter
\@addtoreset{equation}{section}
\renewcommand{\theequation}{\thesection.\@arabic\c@equation}
\makeatother

\makeatletter
\renewcommand\appendix{\par
  \setcounter{section}{0}%
  \setcounter{subsection}{0}%
  \gdef\thesection{Appendix \@Alph\c@section }
  \renewcommand{\theequation}
  {\Alph{section}.\arabic{equation}}
}
\makeatother

\newcommand{\ba}{\begin{eqnarray}}
\newcommand{\ea}{\end{eqnarray}}

\date{}
\begin{document}

\begin{titlepage}

\vspace*{-15mm}   
\baselineskip 10pt   
\begin{flushright}   
\begin{tabular}{r}    
{\tt CCQCN-2015-89} 
\\
{\tt CCTP-2015-12}
\\   
{\tt SU-ITP-15/06}
\\   
{\tt YITP-15-40}
\\   
May 2015
\end{tabular}   
\end{flushright}   
\baselineskip 24pt   
\vglue 10mm

\begin{center}
  {\LARGE
  
Instanton dynamics in finite temperature QCD\\ via holography  
  }
\end{center}
\vspace{0.2cm}
\baselineskip 18pt 
\renewcommand{\thefootnote}{\fnsymbol{footnote}}
\setcounter{footnote}{1}

\begin{center}
Masanori H{\sc anada}${}^{a,b,c}$\footnote{%
E-mail address: hanada@yukawa.kyoto-u.ac.jp}, 
Yoshinori~M{\sc atsuo}${}^{d,e}$\footnote{%
E-mail address: matsuo@physics.uoc.gr
}, and  
Takeshi {\sc Morita}${}^{f}$\footnote{%
E-mail address: morita.takeshi@shizuoka.ac.jp
}

\renewcommand{\thefootnote}{\arabic{footnote}}
\setcounter{footnote}{0}

\vspace{0.4cm}

{\small\it

${}^a$ Yukawa Institute for Theoretical Physics, Kyoto University, \\
Kitashirakawa Oiwakecho, Sakyo-ku, Kyoto 606-8502, Japan.

${}^b$ The Hakubi Center for Advanced Research, Kyoto University, \\
Yoshida-Ushinomiya-cho, Sakyo-ku, Kyoto 606-8501, Japan.

${}^c$ Stanford Institute for Theoretical Physics, Department of Physics, \\
Stanford University, Stanford CA 94305, USA.

${}^d$ Crete Center for Theoretical Physics, Department of Physics, \\
University of Crete, 71003 Heraklion, Greece. 

${}^e$ Crete Center for Quantum Complexity and Nanotechnology, \\ 
Department of Physics, University of Crete, 71003 Heraklion, Greece. 

${}^f$ Department of Physics,
Shizuoka University, \\
836 Ohya, Suruga-ku, Shizuoka 422-8529, Japan

}

\end{center}


\vspace{1.5cm}

\begin{abstract}
We investigate instantons in finite temperature QCD via Witten's holographic QCD.
To study the deconfinement phase, we use the setup proposed in \cite{Mandal:2011ws}.  
We find that the sizes of the instantons are stabilized at certain values both
in the confinement and deconfinement phases.
This agrees with the numerical result in the lattice gauge theory.
Besides we find that the gravity duals of the large and small instantons in the deconfinement phase  have different topologies.
We also argue that the fluctuation of the topological charges is large in confinement phase while it is exponentially suppressed in deconfinement phase, 
and a continuous transition occurs at the Gross-Witten-Wadia (GWW) point.
It would be difficult to observe the counterpart of this transition 
in lattice QCD, since the GWW point in QCD may stay at an unstable branch.

\end{abstract}


\end{titlepage}

\newpage
\baselineskip 18pt
\section{Introduction}

Although instantons are essential ingredients in QCD, it is difficult to understand their dynamics because of the strong coupling nature of the theory. 
Perturbative calculations are justified only for small instantons or at high temperature, 
and a suitable effective theory which describes instantons is not known.
Hence we may have to rely on numerical calculations in lattice gauge theory to illuminate their properties.

One possible tool for analyzing instantons is holographic QCD proposed by Witten \cite{Witten:1998zw}.
Although holographic QCD is different from real QCD in quantitive  details,  
it has successfully explained various qualitative aspects of large-$N$ QCD \cite{Gross:1998gk, Aharony:1999ti, Kruczenski:2003uq, Sakai:2004cn}. (See also \cite{Kim:2012ey, Rebhan:2014rxa} for recent developments.)
Hence we expect that holography can also reveal the natures of instantons.
In particular, we focus on the dynamics of  instantons at finite temperature in this study.

Through holographic QCD, instantons at low temperature (in confinement phase) have been studied in 
\cite{Witten:1998uka, Barbon:1999zp, Bergman:2006xn}.
It was shown that the energy of an instanton 
with a particular size approaches to zero in the large-$N$ limit, 
which indicates the large fluctuations of the topological charge in the confinement phase. 
This is consistent with the previous theoretical insights \cite{Witten:1980sp} and numerical calculations in lattice gauge theory \cite{Lucini:2004yh}.

However, dynamics of instantons in deconfinement phase is less clear. 
Although the perturbative calculations work in certain circumstances \cite{Gross:1980br}, the whole instanton dynamics has not been understood.
Their dynamics around the critical temperature would be particularly important 
to reveal the mechanism of the phase transition, 
and hence it is interesting to study it in holographic QCD.
In this direction, the black D4-brane geometry \cite{Aharony:1999ti}, which was supposed to be the gravity dual of the deconfinement phase, 
has been studied initially.
In particular, some agreements 
with the expected properties of the instantons were reported in Ref.~\cite{Bergman:2006xn}.
However there are also some disagreements.
For example, the instanton density $n(\rho,T)$, 
which is the vacuum expectation value (vev) of the single QCD instanton with a size $\rho$ at temperature $T$, shows unexpected behaviours. 
In holographic QCD, the instanton density is calculated from the DBI action of a D0-brane \cite{Witten:1998uka, Barbon:1999zp, Bergman:2006xn}, and the result in the black D4-brane background is given by \cite{Bergman:2006xn}
\begin{align}
n(\rho,T) \propto e^{-S_{\text{DBI}}} &= 
\begin{cases}
\displaystyle \exp\left(- \frac{ 8\pi^2 N }{  \lambda_{YM} } \right)
&\qquad \left(\rho \lesssim 1/T \right) \\
0 &\qquad \left(\rho \gtrsim 1/T \right)
\end{cases}
\label{DBI-D0}
\end{align}
where $\lambda_{YM} $ is a dimensionless 't Hooft coupling 
which we will define below equation (\ref{metric-SD4}).
Thus it does not depend on either $\rho$ or $T$ if $\rho \lesssim 1/T$, and the size of the instanton is a moduli in this region.

However, both perturbative QCD and numerical calculation in lattice gauge theory predict different results.
Perturbative QCD  predicts that the instanton density for a small instanton at $T=0$ is 
\begin{align}
 n(\rho,0) \propto \exp \left(- \frac{8 \pi^2}{g^2(\rho)}   \right) 
  \propto  \rho^{ \frac{11N}{3}-5},
  \label{instanton-QCD-small}
\end{align}
 where $g^2(\rho)$ is the coupling at scale $\rho$ \cite{Gross:1980br}.
 Thus small instantons are suppressed.
At high temperature $T \gg T_c$ in the deconfinement phase,  because of the electric screening,  large instantons would be suppressed.
Indeed the perturbative calculation shows the large instanton suppression as
\begin{align}
 n(\rho,T) = n(\rho,0)  \exp \left(
-\frac{2N}{3} \left(\pi \rho T \right)^2
 -\log\left(1+ \frac{1}{3}(\rho T)^2 \right) \right)
 \label{instanton-QCD-large}
\end{align}
 for $\pi \rho T \gg 1$ and $g^2(T)\ll 1$  \cite{Gross:1980br}.
Although the perturbative calculations are valid only in limited parameter regimes, 
such suppressions of small and large instantons 
would hold for any temperature in the deconfinement phase, 
and then 
the instanton density would have a peak at a finite value.
Actually this tendency has been observed in lattice calculations in the deconfinement phase \cite{Lucini:2004yh}.

These results clearly disagree with the holographic result (\ref{DBI-D0}) in the black D4-brane geometry.
Although holographic QCD cannot reproduce the actual QCD results quantitatively in principle \cite{Witten:1998zw}, qualitative aspects of QCD are expected to be captured. 
Hence this discrepancy 
is a serious puzzle in holographic QCD.
More recently, it has been argued that the black D4-brane geometry  cannot be identified with the deconfinement phase in four-dimensional QCD; 
rather, a geometry called localized solitonic D3-brane will correspond to the deconfinement phase in QCD \cite{Mandal:2011ws}. 
In this article, we study the instantons in this new setup, and see that the instanton density obtained from the localized D3-brane geometry 
satisfies the expectations from QCD.

We find that the size distribution of the instantons is peaked at a finite value, and 
becomes delta-functional at large-$N$.
Interestingly, the topology of the gravity dual of the stable instantons differs from that of the small ones.
Also, we will see that fluctuations of the topological charge, which is large in the confinement phase and suppressed in the deconfinement phase, 
would smoothly change at the Gross-Witten-Wadia type (GWW) point \cite{Gross:1980he, Wadia:1980cp, Wadia:2012fr}.

This paper is organized as follows. 
We begin in section~\ref{sec-review} by reviewing the holographic QCD at finite temperature and discussing the geometries corresponding to the confinement and deconfinement phases. 
Then in section~\ref{sec-confinement}, we argue instantons in the confinement geometry.
These two sections are mostly a review of the previous studies.
In section~\ref{sec-deconfinement}, we argue instantons in the deconfinement phase.
We also argue the $\theta$ dependence and topological susceptibility in section~\ref{sec-theta}, and show that a continuous transition of the susceptibility occur at the GWW point in section~\ref{sec-GWW}.

\section{Confinement and deconfinement phase in holographic QCD}
\label{sec-review}

In this section we review 
the confinement and deconfinement phases in four-dimensional
$SU(N)$ pure Yang-Mills theory in Witten's holographic QCD model \cite{Witten:1998zw}. 
Let us first consider a ten-dimensional Euclidean spacetime, whose $x_4$-direction is compactified 
on a circle with period $L_4$, which we call $S^1_{L_4}$.   
We consider $N$ D4-branes wrapping on this circle. 
For the fermions on the branes, we take the anti-periodic boundary condition along $S^1_{L_4}$ 
so that supersymmetry is broken.

By taking the large $N$ limit of this system {\em a la} Maldacena at low temperature \cite{Maldacena:1997re, Itzhaki:1998dd}, we obtain the dual gravity description of the compactified five-dimensional SYM theory on the D4-branes \cite{Witten:1998zw}, which consists, at low temperature, of a solitonic D4-brane solution wrapping the
$S^1_{L_4}$.
The explicit metric and dilaton is given by \cite{Itzhaki:1998dd}
\begin{align}
ds^2 &= \alpha' \left[\frac{u^{3/2}}{\sqrt{ \lambda_5/4\pi }}
\left( dt^2  + \sum_{i=1}^{3} 
dx_i^2+f_4(u) dx_4^2 \right)+ \frac{\sqrt{ \lambda_5/4\pi }}{u^{3/2}}\left(  
\frac{du^2}{f_4(u)} 
+ u^2 d\Omega_{4}^2 \right)  \right], \nonumber \\
&\quad f_4(u)=1-\left( \frac{u_0}{u}\right)^3 , \quad e^{\phi}=\frac{\lambda_5}{(2\pi)^2N}
\left(\frac{u^{3/2}}{\sqrt{\lambda_5/4\pi}}  \right)^{1/2}  .
 \label{metric-SD4}
\end{align}
This solution also has a non-trivial 
five form potential which we do not show explicitly. 
Here $\lambda_{p+1}$ is the 't Hooft coupling on the D$p$-brane world-volume, which is 
given in terms of the string coupling $g_s$ and Regge parameter $\alpha'$ as 
$\lambda_{p+1} = (2\pi)^{p-2} g_s \alpha'^{(p-3)/2}N $. 
We will also use the dimensionless coupling $\lambda_{YM} \equiv 2\lambda_5/L_4$.

Since the $x_4$-cycle shrinks to zero at $u=u_0$, in order to avoid
possible conical singularities we must choose the asymptotic periodicity
$L_4$ as 
\begin{align}
\frac{L_4}{2\pi}= \frac{\sqrt{\lambda_{5}/4\pi}}{3} u_0^{-1/2} .
\label{u0-beta}
\end{align} 
With this choice, the contractible $x_4$-cycle, together with the
radial direction $u$, forms a so-called cigar geometry, which is
topologically a disc. 
Note that this gravity solution is reliable in the regime $\lambda_{YM} \gg 1$ where the stringy corrections are suppressed.

Witten pointed out that four-dimensional pure Yang-Mills theory is obtained in a regime $\lambda_{YM} \ll 1$, 
because the KK modes about $S^1_{L_4}$ and matter fields (fermions and adjoint scalars which acquire masses via loop corrections) 
in the five-dimensional super Yang-Mills theory are decoupled.
Although this QCD regime ($\lambda_{YM} \ll 1$) and the strong coupling regime ($\lambda_{YM} \gg 1$), where the gravity analysis is reliable, are completely opposite, their properties would be qualitatively related as far as no transition occurs between them.
(This is analogous to the strong coupling expansion of the lattice gauge theory.)
Indeed there is a lot of evidence which supports this connection, and we expect that supergravity analyses capture qualitative aspects of large-$N$ QCD.

So far, we have considered four-dimensional pure Yang-Mills theory at zero temperature. 
In order to study properties at finite temperature, 
we compactify the Euclidean temporal dimension to a circle, and identify its circumference $\beta$ with the inverse temperature, 
$\beta=1/T$. 
In four-dimensional theories with fermions, the anti-periodic boundary condition along the temporal circle is imposed for fermions. 
In Witten's setup, however, fermions in five-dimensional theories decouple in the four-dimensional limit $\lambda_{YM} \to 0$, 
and hence we do not have to impose the anti-periodic boundary condition. 
Rather, ref.~\cite{Mandal:2011ws} argued that the periodic boundary condition is more useful 
to investigate the QCD deconfinement phase via supergravity.

As we decrease $\beta=1/T$ in the geometry (\ref{metric-SD4}), it reaches $\mathcal O\left(L_4/\sqrt{\lambda_{YM}}\right)$, which is the order of the effective string length at $u=u_0$ \cite{Gross:1998gk, Aharony:1999ti}.
Below this value, winding modes of the string wrapping on the $\beta$-cycle could be excited.
Thus the gravity description given
by (\ref{metric-SD4}) would be valid only if
\begin{align} 
\beta \gg \frac{L_4}{\sqrt{\lambda_{YM}}}.
\label{winding-SD4}
\end{align}
In order to avoid this problem,
we perform the T-duality transformation along the $t$-cycle and go to the IIB
frame, where  the solitonic D4 solution becomes solitonic D3-brane solution 
uniformly smeared on the dual $t$-cycle.%
\footnote{Note that the T-duality along the $t$-cycle maps the system to the IIB string theory since the periodicity of the fermions along this cycle is taken to be periodic. If we took the anti-periodic boundary condition, the system is mapped to the 0B string theory 
in which the brane solution has not been studied well.
Hence Ref.~\cite{Mandal:2011ws} took the periodic boundary condition and focused on the IIB supergravity.
 However, it may be possible to derive similar results from the 0B theory too.
 Another difference in the case with the anti-periodic boundary condition 
 is the existence of the black D4-brane solution which is stable for $T>1/L_4$ at strong coupling ($\lambda_{YM} \gg \beta/L_4$). 
 (Note that the black D4-brane solution is not allowed if we took the periodic condition.)
Although this solution is thermodynamically favoured at strong coupling, this solution is not related to the four-dimensional QCD \cite{Mandal:2011ws, Aharony:2006da}. (Roughly speaking, this solution is an analogue of the ``doubler'' in the lattice gauge theory.) 
Indeed this solution is not stable in the weak coupling where we obtain the QCD \cite{Mandal:2011ws, Aharony:2006da}, and we should remove this solution by hand if we study the QCD through the holographic QCD with the anti-periodic boundary condition.
} 
From now, $t'$ and $\beta'$ denote the dual temporal coordinate and its period  
\begin{align}
\beta'\equiv \frac{(2\pi)^2}{\beta}=(2\pi)^2T.
\end{align}
In this frame, the mass of the winding strings become heavier as $\beta$ decreases (hence the dual radius $\beta'$ increases)
as opposed to those in the IIA frame, and we can explore the model at higher temperature.

\begin{figure}
\begin{center}
\includegraphics[scale=0.90]{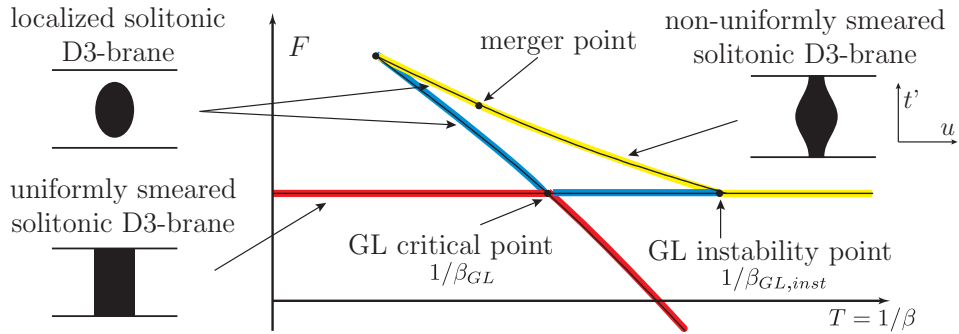}
\caption{The schematic relations among the free energies of various
  solutions in the GL transition and their topologies. 
   The red lines denote the stable solutions.  The blue
  lines denote the meta-stable solutions.  The yellow lines denote the
  unstable solutions.  The solitonic D4-brane (uniformly smeared solitonic
  D3) becomes unstable at the GL instability point.
  The non-uniformly smeared solitonic D3-brane solution appears at this GL instability point and merges to the localized solitonic D3-brane solution at the merger point. 
   }
\label{fig-free_energy}
\end{center}
\end{figure}

As $\beta$ decreases, the uniformly smeared solitonic D3-brane  solution becomes
thermodynamically unstable at a certain temperature, which is called the Gregory-Laflamme (GL) instability point \cite{Gregory:1994bj},%
\footnote{
GL instabilities have been studying in black strings which are the double Wick rotation ($t' \leftrightarrow x_4$) of the smeared soliton. As far as thermodynamical properties, we can read off the soliton results from the black hole ones.
} and is numerically given by
\cite{Harmark:2004ws}
\begin{align} 
\beta_{GL,inst} \simeq  14.4 \frac{L_4}{\lambda_{YM}}.
\label{GL-point}
\end{align} 
See also Figure \ref{fig-free_energy}.
It is expected that, even before $\beta$ is lowered all the way down to
$\beta_{GL,inst}$, the smeared solitonic D3-brane solution becomes meta-stable and
undergoes a first order Gregory-Laflamme (GL) transition at an inverse temperature $\beta_{GL}$ which is approximately given by \cite{Mandal:2011ws}
\begin{align}
\beta_{GL} \sim \left(\frac{3}{2} \right)^7 \frac{
  L_4}{\lambda_{YM}}=17.1\frac{
  L_4}{\lambda_{YM}}, 
\label{GL-t}
\end{align}
 leading to a more stable configuration of D3-branes 
localized on the dual cycle, whose topology is different from the smeared D3-brane solution. 
See Figure \ref{fig-free_energy}.
This figure also shows that the localized solitonic D3-brane solution ceases to exist if $\beta$ is too large; intuitively, if $\beta$ were too large, the
dual cycle would become smaller than the size of the localized soliton, which is not possible. 
The metric of the localized solitonic D3-brane geometry is approximately given by that of 
D3-branes on $R_{9}\times S^1_{L_4}$ for a sufficiently large radius $\beta'$ of the dual
cycle \cite{Harmark:2004ws, Harmark:2002tr}, which we will see in section \ref{sec-LSD3}.

It is argued in \cite{Mandal:2011ws} that this localized solitonic D3-brane geometry can naturally be regarded as a counterpart of 
the deconfinement phase in QCD, and the confinement/deconfinement transition can be identified with 
the GL transition with the transition temperature%
\footnote{Indeed there are various evidences which show the resemblance   
between the GL transition and the confinement/deconfinement transition.
We can show that the phase transition in the five-dimensional SYM theory at strong coupling indeed occurs around (\ref{GL-t}) by applying the analysis in \cite{Morita:2014ypa}.
Several calculations in low dimensional gauge theories also agree with this proposal \cite{Aharony:2004ig, Kawahara:2007fn, Mandal:2009vz, Catterall:2010fx}.
Besides \cite{Hanada:2007wn, Azeyanagi:2009zf, Azuma:2012uc, Azuma:2014cfa} revealed that the confinement/deconfinement transitions exhibit similar properties to the GL transitions.}
\begin{align}
T=\frac{1}{\beta_{GL}} \sim T_c \equiv \left( \frac{2}{3} \right)^7 \frac{\lambda_{YM}}{L_4}.
\end{align}

In addition to the uniformly smeared D3-brane and localized D3-brane solutions, there is another solution:
solitonic D3-brane non-uniformly smeared on the $t'$-cycle.  
This solution describes the D3-branes localized on the $t'$-circle but there is no gap.
(See Figure~\ref{fig-free_energy}.)
Thus this has the same topology to the uniformly smeared solitonic D3-brane geometry while the translation symmetry is broken.
The metric of this solution is perturbatively derived in \cite{Harmark:2004ws}.  
This non-uniform solution arises at the GL instability point. 
 Although the behavior of the non-uniform solution has not fully understood, it is expected that this solution 
merges with the localized solitonic D3-bran solution as shown in
Figure~\ref{fig-free_energy} 
\cite{Aharony:2004ig, Kudoh:2004hs, Kol:2004ww, Sorkin:2006wp, Harmark:2007md}.
This point is called ``merger point.'' 

What is the corresponding phase to the non-uniform solution in QCD? 
Recall that we have taken the T-duality along $t$-circle and the T-duality maps the locations of the branes to the eigenvalues of the Polyakov loop operator $ \exp\left( i \oint_\beta A_0 \right)$.
Thus the non-uniform D3-brane solution describes a phase characterized by the non-uniform eigenvalue distribution of the Polyakov loop operator.
Indeed such a phase is well known in large-$N$ gauge theories although it may be unstable \cite{Semenoff:1996xg, Aharony:2005bq, AlvarezGaume:2005fv}. 
In particular, the merger point is an analogue of the GWW point in the two-dimensional Lattice gauge theory  \cite{Gross:1980he, Wadia:1980cp, Wadia:2012fr}. 
In section \ref{sec-GWW}, we will discuss the importance of the merger point for understanding how the difference of the topological fluctuations at low and high temperatures arises.

\section{Instanton in confinement phase}
\label{sec-confinement}
\begin{figure}
\begin{center}
\includegraphics[scale=.50]{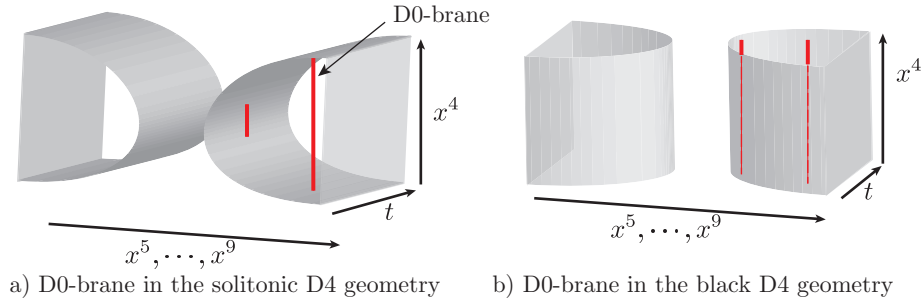}
\caption{Brane configurations of the QCD instanton: (a) D0-brane in the solitonic D4-brane geometry, 
which describes the confinement phase, and (b) D0-brane in the black D4-brane geometry, although it does not correspond to any QCD phases.
In the confinement case, the D0-brane can shrink to a point at the tip of the soliton and the DBI action becomes zero. 
In the black D4-brane background, the effective length of the D0-brane does not depend 
on the position, and hence, the potential is constant as in equation (\ref{DBI-D0}).}
\label{fig-instanton}
\end{center}
\end{figure}

Now we consider instantons. 
First we review the instantons in confinement phase in holographic QCD.
In the bulk theory, the QCD instanton corresponds to the D0-brane winding on the $x_4$-circle  \cite{Witten:1998uka, Barbon:1999zp, Bergman:2006xn}. 
Then the brane configuration of this system is summarized as 
\begin{eqnarray}
\begin{array}{lcccccccccc}
& (0) & 1 & 2 & 3 & (4) & 5 & 6 & 7 & 8 & 9 \\
N~\text{D4-branes} & - & - & - & - & - &&&&& \\
\text{D0-brane (QCD instanton)}
&  &  &  &  & -  &  &  &  & &  
\end{array}
\label{brane-config-IIA}
\end{eqnarray}
Here the parentheses denote the compact directions.
To investigate the potential for an instanton 
we evaluate the DBI action of the single D0-brane.
Ref.~\cite{Barbon:1999zp, Bergman:2006xn} showed that 
the DBI action in the solitonic D4-brane geometry (\ref{metric-SD4}), 
which will correspond to the confinement phase, is
\begin{align}
S_{\text{D0}}=\frac{8\pi^2 N}{\lambda_{YM}}\sqrt{1-\frac{u_0^3}{u^3}}
\label{DBI-D0-SD4}
\end{align}
where $u$ is the position of the D0-brane along the radial coordinate.
Therefore the D0-brane is attracted toward the horizon of the D4-soliton ($u=u_0$), and the classical action disappears when it arrives at the horizon.%
\footnote{If we evaluate the backreaction of the D0-brane, we see that the energy of the D0-brane is not exactly zero.
The solitonic D4-brane solution with non-zero D0-brane charge (D0-D4 geometry) has been calculated in \cite{Barbon:1999zp}, and the D0-branes cost the energy $ \sim V_3 L_4^7 n_0^2/\lambda_5^3$, where $n_0 \equiv N_0/ \beta V_3$ is the charge density for $N_0$ D0-branes and $V_3$ is the spatial volume of $R^3$ in QCD. 
Here we have assumed that the density $n_0$ is small and uniform on the four-dimensional space.
See also \cite{Wu:2013zxa, Seki:2013nta} for the application of the D0-D4 geometry to holographic QCD.
} 
This result can intuitively be understood through the cartoon of the brane configuration depicted in Figure~\ref{fig-instanton}~(a).
Since the D0-brane can shrink to a point at the tip of the soliton, 
the DBI action becomes zero there.

Now we interpret this result as the corresponding QCD instanton dynamics.
The position $u$ of the D0-brane would be related to the size $\rho$ of the instanton \cite{Barbon:1999zp, Bergman:2006xn}.
In the case of the extremal D$p$-brane geometries, this relation can be understood explicitly.
Roughly speaking, in these geometries, the typical energy scale at the radial position 
$u$ is given by $u^{(5-p)/2}/\sqrt{\lambda_p}$ where $\lambda_p$ is the 'tHooft coupling on the D$p$-branes  \cite{Itzhaki:1998dd, Susskind:1998dq, Peet:1998wn} and hence the size of the instanton is related to its inverse $\sqrt{\lambda_p}/u^{(5-p)/2}$.  
Indeed this relation has been confirmed in the AdS$_5$/CFT$_4$ case \cite{Aharony:1999ti, Chu:1998in, Kogan:1998re, Bianchi:1998nk, Balasubramanian:1998de}. 
However this argument cannot be applied to the confinement geometry (\ref{metric-SD4}) at finite temperature, since there are two energy scales $ \sqrt{u/L_4 \lambda_{YM}}$ and $ \sqrt{u_0/L_4 \lambda_{YM}}$.%
\footnote{
\label{ftnt-instanton-size}
In the AdS$_5$/CFT$_4$ correspondence \cite{Gubser:1998bc, Witten:1998qj}, we read off  $\langle \text{Tr}F \wedge F\rangle$ in the gauge theory from the boundary value of the RR scalar field which is sourced by a D-instanton in the bulk \cite{Aharony:1999ti, Chu:1998in, Kogan:1998re, Bianchi:1998nk, Balasubramanian:1998de}. 
Hence the wave equation of the scalar in the AdS$_5$ fixes the instanton dynamics on the boundary. 
Importantly the wave equation for the S-wave can be rescaled so that it is described by a single dimensionless parameter 
$\lambda_3 k^2 / u^{2}$ where $k$ is the longitudinal momentum \cite{Peet:1998wn}.
It leads us to the scaling behavior $\rho \sim 1/k \sim \sqrt{\lambda_3}/ u$ of the size $\rho$ of the instanton corresponding to the D-instanton located at $u$. 
Similar scaling, with the dimensionless parameter $\lambda_p k^2 / u^{(5-p)}$, 
would be obtained in other extremal D$p$-brane cases too.
However the metric of the confinement geometry (\ref{metric-SD4}) involves the additional factor $f=1-(u_0/u)^3$ and the scaled wave equation depends on both $\lambda_4 k^2 / u$ and $\lambda_4 k^2 / u_0$. 
Thus we obtain two energy scales $ \sqrt{u/L_4 \lambda_{YM}}$ and $ \sqrt{u_0/L_4 \lambda_{YM}}$.
} 
(Through (\ref{u0-beta}), the latter becomes $ \sqrt{u_0/L_4 \lambda_{YM}} \sim 1/L_4$, which is the same order to the glueball masses in the holographic QCD \cite{Gross:1998gk}.)
Although we do not have explicit relation between $\rho$ and $u$ in the confinement geometry, we naively assume \cite{Barbon:1999zp, Bergman:2006xn}
\begin{align}
\rho \sim \sqrt{\frac{L_4 \lambda_{YM}}{u}}.
\label{instanton-size-u}
\end{align}
This assumption would be valid at least when $u \gg u_0$ where $u_0$ would be irrelevant or when $u \sim u_0$ where the two energy scales are coincident.
Once we admit this assumption, the equation (\ref{DBI-D0-SD4}) indicates a suppression of a small instanton in the confinement phase, 
which is qualitatively consistent with the perturbative QCD (\ref{instanton-QCD-small}).
At $u\sim u_0$ ($\rho \sim \sqrt{L_4 \lambda_{YM}/u_0} \sim L_4$), 
the instanton can exist with the zero value of the DBI action, 
which would imply that the fluctuation of the topological charge is large \cite{Witten:1998uka}. 
Moreover a larger instanton cannot exist.
Thus the instanton density has a sharp peak at 
\begin{align}
\rho_{\text{peak}} \sim L_4
\label{rho-peak}
\end{align}
 for large $N$.
The location of the peak does not depend on temperature. 
This would be related to the large-$N$ volume independence \cite{Eguchi:1982nm, Gocksch:1982en}.  
Remarkably the lattice calculation in the confinement phase yields a similar sharp and temperature independent peak in the instanton density \cite{Lucini:2004yh}. 

\section{Instanton in deconfinement phase}
\label{sec-deconfinement}

To investigate the thermodynamics of QCD at high temperature through holography, we need to take the T-dual along the Euclidean time circle as argued in Section \ref{sec-review}. Then the brane configuration (\ref{brane-config-IIA}) is mapped to the IIB frame: 
\begin{eqnarray}
\begin{array}{lcccccccccc}
& (0') & 1 & 2 & 3 & (4) & 5 & 6 & 7 & 8 & 9 \\
N~ \text{D3-branes} &  & - & - & - & - &&&&& \\
\text{D1-brane (QCD instanton)}
& - &  &  &  & -  &  &  &  & &  
\end{array}
\label{brane-config-IIB}
\end{eqnarray}
Thus we should consider a D1-brane instead of a D0-brane to study the dynamics of the QCD instanton.

\begin{figure}
\begin{center}
\includegraphics[scale=.50]{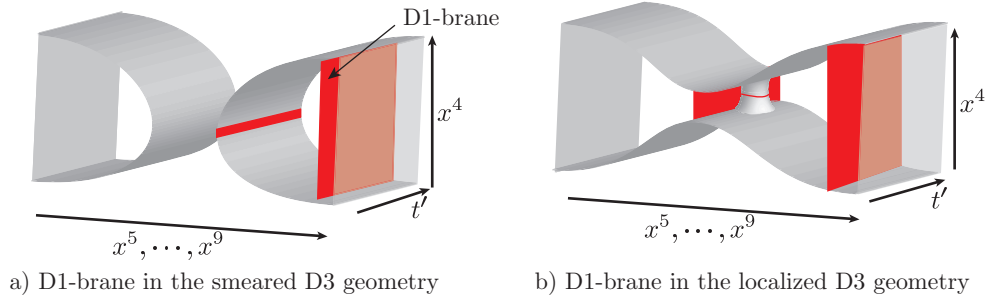}
\caption{Brane configurations of the QCD instanton in the T-dual picture: 
(a) the confinement phase (D1-brane in the smeared D3-brane geometry) and (b) the deconfinement phase (D1-brane in the localized D3-brane geometry). 
In the confinement case, the D1-brane can shrink as the D0-brane in the solitonic D4-brane geometry. 
In the deconfinement case, the D1-brane remains a finite volume and the DBI action is always non-zero. }
\label{fig-D3-D1}
\end{center}
\end{figure}

At $T < T_c$, the stable geometry in the IIB frame is the uniformly smeared D3-branes, which is the T-dual of the solitonic D4-brane geometry (\ref{metric-SD4}), and the instanton is described by a D1-brane on this geometry. (See Fig.~\ref{fig-D3-D1} (a).)
Since the T-duality retains the values of the classical action, the results in the confinement phase discussed in the previous section remain the same. 

At $T > T_c$, the stable geometry is given by the localized D3-branes. We study the dynamics of the D1-brane on this geometry by using the probe approximation.

\subsection{Geometry of localized D3-branes on a circle}
\label{sec-LSD3}

First we explain the details of the localized solitonic D3-brane solution, 
which corresponds to the deconfinement phase.
In this geometry, the D3-branes are localized on $t'$-cycle, 
where $t'$ is the euclidean time direction in the IIB frame. 
We set the location of their center of mass to be $t'= 0$.
Then because of the periodicity $t'=t'+ \beta'$ ($\beta'=4\pi^2T $), 
their mirror images sit around $t'=n \beta'$ ($n=\pm 1, \pm 2, \cdots$).
Since the D3-branes and their mirrors are gravitationally interacting, each ``horizon'' is  stretched along $t'$-direction, 
and the spherical symmetry is broken.
Because it is difficult to treat this effect exactly, 
we use an approximation which is justified at $T/T_{GL}\gg 1$, where the interaction becomes weak. 
In particular, at the leading order of this expansion, we can treat the horizon spherically symmetric, and we do not consider higher order corrections in this paper for simplicity.%
\footnote{Although this approximation is valid only for high temperature  $T/T_c \gg 1$, the qualitative properties of the localized D3-brane would not be changed even around $T_c$ as indicated in the numerical calculation of the localized black holes \cite{Kudoh:2004hs}.} 

Although such localized solitonic solutions have not been investigated well, 
localized black brane geometries, which are just the double Wick rotation of the solitonic brane geometries, 
have been studied very actively in the context of the Gregory-Laflamme instability, 
and we can borrow the results.
We consider the wick rotation of the black D3-brane 
localized on $S^1$-circle \cite{Harmark:2004ws, Harmark:2002tr}, 
and then, the metric takes the following form 
\begin{equation}
 ds^2
 =
 H^{-1/2}
 \left[
  \sum_{i=0}^3 dx_i^2 + fdx_4^2
 \right]
 + H^{1/2}
 \left[ 
  \frac{A}{f}dR^2 + \frac{A}{K^{d-2}}dv^2 + KR^2 d \Omega_4^2
 \right] \ ,     
\end{equation}
where
\begin{equation}
 f = 1 - \frac{R_0^3}{R^3} \ , 
\end{equation}
and $A$ and $K$ are functions of $R$ and $v$. 
Here, $x_4$ is not the euclidean time direction, 
but that for $S^1_{L_4}$ as in the IIA frame. 
The euclidean time coordinate, $t'$, 
is included in the $(R,v)$-plane. 

In order to simplify the analysis, we consider two regions: asymptotic region and near region 
\cite{Harmark:2004ws, Harmark:2002tr, Gorbonos:2004uc}.%
\footnote{
In this paper, we use the approximated form for localized neutral black holes in \cite{Gorbonos:2004uc}. 
It is straightforward to obtain the geometry for D3-branes from that for the neutral black holes 
\cite{Harmark:2004ws, Harmark:2002tr}. 
} 
In the asymptotic region (i.e. $u\gg u_H$ or $t'\gg u_H$), effects of the black hole can be calculated by solving 
linearized equations, and then, the metric is given by 
\begin{align}
ds^2 &= \alpha' \left[H^{-1/2}
\left( \sum_{i=1}^{3}  
dx_i^2 + (1+2\Phi)dx_4^2  \right)
 + H^{1/2} (1-\frac{1}{2}\Phi)\left(  du^2 +dt'^2+ u^2 d\Omega_{4}^2  \right)  \right]  , \nonumber \\ 
&
H= \sum_n \frac{2 \lambda_5/\beta }{(u^{2}+(t'-n \beta')^{2})^2}, 
 \quad e^{\phi}=\frac{\lambda_5}{2\pi N \beta}, \nonumber \\
& \Phi=- \frac{u^4_H}{2} \sum_n \left( \frac{1}{u^2+(t'-n \beta')^2} \right)^2, \quad 
 u_H= \sqrt{2 \lambda_5 T}\frac{\pi}{2 L_4}, \quad 
 \beta' = \frac{(2\pi)^2}{\beta}=(2\pi)^2T \ .  
\label{metric-LD3-large-u}
\end{align}
Note that mirrors contribute to the metric. 
In the near region (i.e. $u,t' \sim u_H$), the effect of the black hole becomes much larger than 
that of the mirror images. 
In this limit, the metric is given by 
\begin{align}
ds^2 &= 
 \alpha' \left[
 H^{-1/2} \left( \sum_{i=1}^{3} dx_i^2 
 + \left(\frac{1-\frac{r_0^4}{r^4}}{1+\frac{r_0^4}{r^4}}\right)^2 dx_4^2  \right)
 + H^{1/2} \left(1+\frac{r_0^4}{r^4}\right) \left(d r^2 + r^2 d \Omega_5^2\right) 
 \right] \ , 
\notag\\&
 H = \frac{2 \lambda_5/\beta}{r^4}\left(1+\frac{r_0^4}{r^4}\right)^{-2}\ , \quad
 r_0 = \frac{u_H}{\sqrt 2} = \frac{\pi\sqrt{\lambda_5 T}}{2 L_4} \ , \quad e^{\phi}=\frac{\lambda_5}{2\pi N \beta},
\label{metric-near-region}
\end{align}
where we have defined the coordinate $r$ such that it 
approaches to the Newtonian gauge as $r\gg r_0$. 

In the following, we introduce a midpoint $r_1$, and treat $r<r_1$ and $r>r_1$ as near and asymptotic regions, respectively. 
\subsection{The D1-brane in the asymptotic region}
\label{sec:D1-asympt}

We first consider a D1-brane located in the asymptotic region 
where the background metric is approximated by \eqref{metric-LD3-large-u}. 
By regarding the brane configuration (\ref{brane-config-IIB}) and the symmetry of the background geometry, the D1-brane will be embedded in $t'$, $x_4$ and $u$ space. 
We take $(t', x_4)$ as the world volume coordinates on the D1-brane, and then the induced metric is given by 
\begin{equation}
 ds^2_\text{D1} = \alpha' \left[H^{-1/2} (1+2\Phi)dx_4^2
 + H^{1/2} \left(1-\frac{1}{2}\Phi\right)\left(1 + \left(\frac{dU(t')}{dt'}\right)^2 \right)dt^{\prime 2}  \right]  \ .
 \label{induced-D1}
\end{equation}
Then the DBI action in the asymptotic region becomes
\begin{align}
S_{\text{D1}} &=  \frac{1}{(2\pi) \alpha' } \int_0^{\frac{(2\pi)^2}{\beta}} dt' \int_0^{L_4} dx_4 e^{-\phi} \sqrt{\det g_{\text{D1}}}, \nonumber \\
 &=  \frac{N \beta L_4}{  \lambda_5 } \int_0^{\frac{(2\pi)^2}{\beta}} dt'   \left( 1 + \frac{3}{4}\Phi\right) \sqrt{ 1 +\left( \frac{ dU(t')}{dt'}\right)^2 } . 
 \qquad (u \gg \beta'\gg u_H) 
 \label{dbi-asympt}
\end{align}
Since $t'$ dependence of $\Phi$ can be neglected for large $u$, we approximate that $U(t')$ is constant.
Then we obtain
\begin{align}
S_{\text{D1}}  = \frac{8 \pi^2 N }{  \lambda_{YM} } - N \frac{3 \pi^5   }{2^6} \frac{\lambda_{YM}T}{u^3 L_4^2} + \cdots 
\label{potential-u-large-u}
\end{align}
The second term indicates that the D1-brane is attracted toward the D3-branes ($u=0$).
This is similar to the confinement geometry case (\ref{DBI-D0-SD4}), although the potential is now proportional to temperature.
Around $T \sim T_c$, this term becomes $\sim N u_0^3/\lambda_{YM} u^3$ 
which is the same order to the attractive potential in the confinement phase (\ref{DBI-D0-SD4}) at large $u$, and it becomes stronger as temperature increases. 

This result is valid only for $u \gg u_H$ and the approximation becomes 
worse as $u$ approaches to $u_H$. 
At $u\sim u_H$, \eqref{potential-u-large-u} behaves as 
\begin{equation}
 S_\text{D1} - \frac{8\pi^2 N}{\lambda_{YM}} \sim -\frac{N}{\lambda_{YM}} \sqrt{ \frac{T_c}{T} } \ , 
\label{potential-u_H}
\end{equation}
and for $u<u_H$ the above discussion will completely break down.

\subsection{The D1-brane near D3-branes}
\label{sec-large-instanton}

Since the D1-brane is attracted toward the D3-branes, 
the D1-brane would be stabilized at  $u=0$ and would stretch 
between the D3-branes and their mirror image along the compact $t'$ circle 
as depicted in (b) of Fig.~\ref{fig-D3-D1}.
In \ref{sec-instanton-stability}, 
we demonstrate that the stable classical solution of the DBI action is given by this configuration indeed. 

Note that the hypersurface at $u=0$ of the localized D3-brane geometry  has a topology of $S^2 \times R^3 \times S^4$, and the stable D1-brane wraps on this $S^2$.
On the other hand,  the D1-brane in the asymptotic region ($u \gg u_H$) winds $t'$- and $x_4$-cycles which compose a topology of $T^2$. Thus the topology of this D1-brane differs from that of the stable D1-brane at $u=0$ which winds $S^2$. 
This is because the stable D1-brane reaches the ``horizon'' of the D3-brane where the $x_4$-direction shrinks. 
It indicates that the D1-brane in the asymptotic region cannot continue to the stable D1-brane at $u=0$ smoothly.
When the D1-brane reaches the ``horizon,'' the topology changes.
We will later see that  the D1-brane in the asymptotic region describes a small instanton, 
and it means that the small instanton does not smoothly continue to the stable instanton.


Let us compute the value of the DBI action for the stable D1-brane at $u=0$.%
\footnote{%
If we could calculate the value of the DBI action for the appropriate configuration of the D1-brane corresponding to the QCD instanton with size $\rho$, we would obtain the potential for $\rho$ as we did for the solitonic D4-brane background in section \ref{sec-confinement}.
Since the localized D3-brane does not have the isometry along $t'$-cycle, 
it is difficult to specify the configuration for a specific size of the instanton. 
For this reason, we calculate the action only for the stable classical solution.  
We will discuss a related issue in section \ref{sec-instanton-density-dec}. }
%
Near the D3-branes or their mirror image, the metric can be approximated by that for the near region (\ref{metric-near-region}). 
However, around the middle between the D3-branes and their mirror, 
the metric cannot be described by that for the near region but 
should be approximated by that for the asymptotic region (\ref{metric-LD3-large-u}). 

To evaluate the DBI action for the stable D1-brane at $u=0$, it is convenient to rewrite the metric \eqref{metric-LD3-large-u} and \eqref{metric-near-region} in the following combined expression;  
\begin{equation}
 ds^2 = 
 \alpha' \left[
 H^{-1/2} \left( \sum_{i=1}^{3} dx_i^2 
 + f_4 dx_4^2  \right)
 + H^{1/2} f_r \left(d r^2 + r^2 d \Omega_5^2\right) 
 \right] \ ,  
 \label{metric-unified}
\end{equation}
where $f_4$, $f_r$ and $H$ are functions of $r$ and one of the angular coordinates of $S^5$, which are related to $u$ and $t'$.
In the asymptotic region, they approach to 
\begin{align}
f_4 = 1 + 2 \Phi \ , \quad 
f_r = 1 - \frac{1}{2}\Phi, \quad
 H = \sum_n \frac{2 \lambda_5/\beta }{(u^{2}+(t'-n \beta')^{2})^2}, 
\end{align}
with $r^2=u^2+t^{'2}$.
In the near region, the geometry has spherical symmetry on $S^5$ 
at the leading order and they become
\begin{align}
f_4 = \left(\frac{1-\frac{r_0^4}{r^4}}{1+\frac{r_0^4}{r^4}}\right)^2, \quad 
f_r = 1+\frac{r_0^4}{r^4}, \quad
 H =  \frac{2 \lambda_5/\beta}{r^4}\left(1+\frac{r_0^4}{r^4}\right)^{-2}, 
\end{align}
as is shown in \eqref{metric-LD3-large-u} and \eqref{metric-near-region}. 

In this metric, $u=0$ corresponds to a fixed direction in $S^5$ and 
 $r$ can be chosen to be identified to $t'$ when $u=0$. 
Then, the induced metric on the D1-brane at $u=0$ is expressed as 
\begin{equation}
 ds^2_\text{D1} = 
 \alpha' \left[
 H^{-1/2} f_4 dx_4^2 
 + H^{1/2} f_r(t') d t^{\prime 2} 
 \right] \ ,    
\end{equation}
and the DBI action is given by 
\begin{equation}
 S_\text{D1} = 2 \frac{N \beta}{ \lambda_5 } \int_{r_0}^{\beta'/2} d t' \int_{0}^{L_4} dx_4 \sqrt{f_4f_r} \ . 
\end{equation}
In order to calculate this action, we introduce a mid-point $r_1$ and divide the $t'$-integration into two parts, that for the asymptotic region and 
that for the near region; 
\begin{equation}
 S_\text{D1} = S_\text{near} + S_\text{asymp}	\ .  
\end{equation}
We will soon see that the final result is independent of $r_1$. 

The integration for the near region can be calculated as 
\begin{align}
 S_\text{near} &= 2 \frac{N \beta L_4}{  \lambda_5 }\int_{r_0}^{r_1} d t'
 \frac{1-\frac{r_0^4}{t^{\prime 4}}}{\sqrt{1+\frac{r_0^4}{t^{\prime 4}}}} 
\notag\\&
 = 2 \frac{N \beta L_4 r_0}{  \lambda_5 } \left(-\sqrt{2} + \frac{r_1}{r_0} 
 + \mathcal O\left(\left(\frac{r_0}{r_1}\right)^3 \right)\right) \ .  
\end{align}
Here, we are assuming $r_0\ll \beta'$, and hence, 
we can take $r_1\gg r_0$. 
In the asymptotic region, we can neglect $\Phi$ since it gives contributions at $\mathcal O(r_0^4)$, 
and hence the DBI action can be calculated as 
\begin{equation}
 S_\text{asymp} = \frac{2 N \beta L_4}{  \lambda_5 } \int_{r_1}^{\beta'/2} d t' +\mathcal O(r_0^3) 
 = \frac{N \beta L_4}{  \lambda_5 }\left(\beta' - 2r_1 + \mathcal O(r_0^3)\right) \ . 
\end{equation}
By summing these two results, we obtain 
\begin{equation}
 S_\text{D1} = \frac{N \beta L_4}{  \lambda_5 } 
 \left(\beta' - 2^{3/2} r_0 + \mathcal O(r_0^3)\right) \ . 
\end{equation}
which does not depend on $r_1$.
By using $\beta'=  (2\pi)^2 T$, we finally obtain
\begin{align}
S_{\text{D1}} =
\frac{ 8\pi^2 N }{  \lambda_{YM} }\left(1- \frac{1}{4\pi } 
\sqrt{\frac{\lambda_{YM}}{L_4T}} \right) 
=
\frac{ 8\pi^2 N }{  \lambda_{YM} }\left(1- \frac{3^{7/2}}{2^{11/2}\pi } 
\sqrt{\frac{T_c}{T}} \right) .
\label{LargeInst}
\end{align}
Thus the DBI action is finite and the topological fluctuation is exponentially suppressed at large-$N$. 
\footnote{
If we use the dilute gas approximation, we obtain $\chi_t \propto e^{-S_{\text{D1}}}$ in the localized D3-brane geometry where $\chi_t$ is the topological susceptibility.
}

Note that the value of the action (\ref{LargeInst}) for the stable D1-brane at $u=0$ is the same order to \eqref{potential-u_H}  for $u\sim u_H$ which is extrapolated from the DBI action (\ref{potential-u-large-u}) for large $u$.
It would indicate that the potential (\ref{potential-u-large-u}) at large $u$ continues to the value (\ref{LargeInst}).
Recall that the topologies of the D1-brane at large $u$ and $u=0$ are different, and the topology change occurs when the D1-brane reaches the ``horizon'' of the soliton.
Since the value of the classical action is related to the area of the D1-brane, 
it would be continuous through the topology change. 
However, its (higher) derivative with respect to some deformation parameters of the D1-brane may not be continuous.

\subsection{Size of instantons in the deconfinement phase}
\label{sec-instanton-density-dec}

We have calculated the DBI action of the D1-branes. 
Now we argue the relation between the size $\rho$ of the QCD instantons 
and the radial location $u$ of the D1-branes, as we have done for the confinement phase in section \ref{sec-confinement}.
The relation is more complicated than that for the confinement geometry (\ref{metric-SD4}),  
since we have taken the T-dual on the temporal circle and the energy in the IIA frame would appear in an unusual manner. 
Furthermore, the metric can analytically be expressed 
only by a couple of the approximated forms for two patches. 

Fortunately the asymptotic metric (\ref{metric-LD3-large-u}) has an approximate isometry along the temporal circle 
if $u$ is sufficiently large, and we can consider the IIA frame by taking the T-dual again. 
There the typical enegy scale is  $ \sqrt{u/L_4 \lambda_{YM}}$ 
for the D1-brane which is located at $u$. 
This is the same scale to that of the confinement geometry, since the localised D3-brane geometry \eqref{metric-LD3-large-u} asymptotically approaches to the smeared D3-brane geometry \cite{Mandal:2011ws}, which is the T-dual of the confinement geometry (\ref{metric-SD4}). 

For the near D3-brane metric (\ref{metric-near-region}), 
the isometry along the temporal circle is broken. 
Hence the T-dual picture in the IIA side is not clear and it is hard to see the relation. 
If we consider only the near region, there typical energy scales would be naively
$r/\sqrt{\lambda_{YM}L_4 T}$ and $r_H/\sqrt{\lambda_{YM}L_4 T} \sim 1/L_4$ for the D1-brane located at $r$. 
However we need to consider the connection to the asymptotic region, where the aforementioned different scalings arise,
to estimate the energy of the gauge theory on the boundary.
 (See footnote \ref{ftnt-instanton-size} and \cite{Peet:1998wn}.)
Furthermore, the D1-brane for the stable configuration at $u=0$ is stretched along $r$ direction, and  it makes the situation more complicated.

However, for small instantons 
the contribution of the near region would be irrelevant and the asymptotic metric (\ref{metric-LD3-large-u}) would dominate for obtaining the relation to the instanton size.
Then it would be possible that the $\rho-u$ relation (\ref{instanton-size-u}) in the confinement geometry holds approximately even in the deconfinement phase.
Under this assumption, the DBI action (\ref{potential-u-large-u}) is rewritten as
\begin{align}
S_{\text{D1}} - \frac{8\pi^2 N }{  \lambda_{YM} } \sim - N   \frac{T T_c^5\rho^6}{\lambda_{YM}^7} + \cdots 
\label{potential-rho}
\end{align}
The value of this action is larger for smaller $\rho$ and it suggests that the small size instanton would be suppressed.

For the stable D1-brane at $u=0$, we cannot estimate the instanton size because of the difficulties mentioned above. 
However the distance between the D1-brane at $u=0$ and the boundary ($u=\infty$) is finite, which implies that the instanton size for the D1-brane at $u=0$ would be finite. 
We presume that this corresponds to the largest instanton in QCD and a larger instanton is not allowed. 
Thus the instanton density would have a sharp peak at this value of $\rho$ at large-$N$.

Although we cannot calculate this largest size, we can estimate the lower bound for this size by using the relation (\ref{instanton-size-u}) in the asymptotic region and substituting $u = u_H$,
\begin{align}
 \rho
 \sim \sqrt{\frac{\lambda_{YM}L_4}{u_H}} 
 \sim \left( \frac{\lambda_{YM}L_4^3}{T} \right)^{\frac14} 
\ .
\label{rho-vev}
\end{align}
Note that this lower bound of the peak size (\ref{rho-vev}) increases 
as $T$ decrease and reaches $\rho \sim L_4$ around $T\sim T_c$, 
which is the same order to the peak size of the instanton in the confinement phase (\ref{rho-peak}).%
\footnote{
Recall that  the value of the DBI action at $u \sim u_H$ (\ref{potential-u_H}) obtained from the potential (\ref{potential-u-large-u}) for large $u$ is the same order to the DBI action at $u=0$  (\ref{LargeInst}).
Thus, 
the size (\ref{rho-vev})  obtained from the relation (\ref{instanton-size-u}) at $u \sim u_H$ provides the order of  the largest size of the instanton (the D1-brane at $u=0$), 
if it has a similar property.}

\section{$\theta$-vacuum and topological susceptibility}
\label{sec-theta}
We have studied the instantons in the deconfinement phase. 
The results show that the instanton density has the sharp peak at a finite instanton size but the  energy at this size is still finite. 
This implies that the topological fluctuation would be suppressed in the deconfinement phase.
On the other hand, the zero energy of the instantons in the confinement phase implies the large fluctuation of the topological charge.
To confirm this picture, we investigate the instanton effects in $\theta$-vacuum and estimate the topological susceptibility $\chi_t$, 
which is defined by the second derivative of the free energy with respect to $\theta$ parameter:
\begin{align}
\chi_t \equiv \frac{d^2 F}{d \theta^2} \ . 
\end{align}
We show that the topological susceptibility is indeed suppressed in the localized D3-brane geometry consistently 
with the finite value of the DBI action\footnote{For lattice studies, see e.g. \cite{Gattringer:2002mr,Berkowitz:2015aua,Kitano:2015fla}.}.

We first recall the topological charge at low temperature \cite{Witten:1998uka,   Bergman:2006xn}. 
The confinement phase corresponds to the solitonic D4-brane geometry, 
and the instanton is described by the (euclidean) D0-brane wrapped on the $x_4$-direction.
Thus the topological charge corresponds to the ``RR-charge'' and 
can be estimated from the configuration of the RR 1-form $C_1$. 
The parameter $\theta$, which is the chemical potential of the instanton, 
corresponds to the boundary condition of $C_1$; 
\begin{equation}
 \theta = \int_{S^1_{L_4}} C_1 \ , 
\end{equation}
where integration is over $S^1_{L_4}$, which is $x_4$-direction at the boundary $u\to\infty$. 
In the case of the solitonic D4-brane geometry, $S^1_{L_4}$ is the boundary of a disk $D$. 
By using the Stokes theorem, it can be written in terms of the field strength $F_2=dC_1$; 
\begin{equation}
 \theta = \int_D F_2 \ . 
\end{equation}
This implies that the field strength is proportional to $\theta$. 
Then, the classical action for the bulk RR-field is estimated as; 
\begin{equation}
 S \sim \int F^2 \propto  \theta^2 \ . 
\end{equation}
Therefore, the free energy has finite quadratic term of $\theta$, 
indicating the topological susceptibility $\chi_t$ is finite. 
Thus the fluctuation of the topological charge is large \cite{Witten:1998uka}.

Now, we turn to the localized D3-brane geometry. 
In this case, the instanton corresponds to the D1-brane 
which is wrapped on the torus $T^2$ of the $(x_4, t')$-plane. 
The parameter $\theta$ is related to the boundary condition of 
the RR 2-form $C_2$; 
\begin{equation}
 \theta = \int_{T^2} C_2 
\end{equation}
where the integration is performed at the boundary $u\to\infty$. 
In order to see the contributions of this boundary condition to the free energy, 
we consider the Stokes theorem; 
\begin{equation}
 \int_{\partial M} C_2 = \int_{M} F_3
\end{equation}
where $F_3$ is the field strength associated to $C_2$ and 
$M$ is the three-dimensional space of $(u, x_4, t')$. 
The $S^1$ circle of $x_4$ shrinks to a point at $r=r_0$, 
which can approximately be expressed in the $(u, t')$-coordinates as 
\begin{equation}
 u^2 + t^{\prime 2} \sim r_0^2 \ . 
\end{equation}
Thus the $x_4$-direction can shrink only at $|t'|\lesssim r_0$. 
The space continues to $u=0$ in the other region, 
and connected to the opposite side of $S^4$. 
Then, the boundary of $M$ consists of two tori, $T^2_+$ and $T^2_-$ at $u=\infty$ with opposite angles, 
and the Stokes theorem provides us with  
\begin{equation}
 \int_M F_3 = \int_{T^2_+} C_2 - \int_{T^2_-} C_2 \ , 
 \label{LD3-Stokes}
\end{equation}
where the minus sign in the last term comes from 
the difference of the orientation for two tori. 
The 2-form field $C_2$ takes the same value on $T^2_+$ and $T^2_-$ 
since they are the opposite points on $S^4$ and 
we assume that the solution has the spherical symmetry. 
This implies cancellation in the r.\ h.\ s., 
and hence, the field strength is not constrained by $\theta$. 
Thus the topological susceptibility $\chi_t$ vanishes and the fluctuation is suppressed as we expected.%
\footnote{
The black D4-brane geometry also shows $\chi_t=0$ \cite{Bergman:2006xn}, 
even though it does not provide the correct instanton density. 
In this case, $(u,x_4)$ plane is terminated at the horizon, and hence 
it has the topology of the cylinder. 
The parameter $\theta$ is given in terms of the RR 1-form by 
\begin{equation}
 \theta = \int_C F_2 - \int_{S^1_H} C_1 \ , 
\end{equation}
where $S^1_H$ is the $x_4$ circle at the horizon. 
Now $F_2$ can be zero by adjusting the second term according to $\theta$,
and then, the topological susceptibility $\chi_t$ is zero. 
We can explain $\chi_t=0$ in a similar fashion even in the case of the localized D3-branes. 
If we restrict $(u, x_4, t')$ plane to $u\geq 0$ by using the spherical symmetry on $S^4$, 
the integration of $C_2$ at $u=0$ appears instead of the last term in \eqref{LD3-Stokes}. 
It can be adjusted such that it cancels the integration of $C_2$ at the boundary. 
}

\section{Continuous transition of topological fluctuation at GWW point}
\label{sec-GWW}

As we have shown in section \ref{sec-deconfinement}, the DBI action of the D1-brane in the localized solitonic D3-brane  geometry remains finite and the topological charge fluctuation is exponentially suppressed even at the critical temperature (\ref{GL-t}).
This is not surprising since the confinement/deconfinement transition (GL transition) is of first order.

As in Fig.~\ref{fig-free_energy}, the localized D3-brane branch is connected to the confinement geometry (uniformly smeared D3-brane) through the non-uniformly smeared D3-brane geometry.
By tracking this, we can see how the physics in the deconfinement phase changes to that in the confinement phase. 
Then an important question is where and how the suppression of the topological fluctuation in the localized D3-brane geometry changes to the large fluctuation in the confinement geometry.
We propose that it will occur at the merger point where the localized solitonic D3-brane   geometry merges to the non-uniformly smeared solitonic D3-brane  geometry and the topology changes. 
Since there is no ``gap'' along $t'$-circle in the non-uniform D3-brane  geometry similar to the uniform D3-brane geometry, the DBI action of the D1-brane in this geometry can be zero.
On the other hand in the localized D3-brane geometry, due to the existence of the gap, the DBI action is finite.
As this gap is becoming smaller, the DBI action will be  smaller and would reach zero at the merger point.
Therefore the continuous transition would occur at the merger point.

This is also consistent with the analysis in the $\theta$-vacuum. 
If the geometry has only single boundary, the field strength of the RR-field is constrained by $\theta$ 
and the topological susceptibility becomes finite. 
In the localized D3-brane geometry, $x_4$-direction does not shrink in a specific region 
and then, the 3-form flux reach to the opposite side of $S^4$. 
This effectively plays the role of different two boundaries. 
However, the ``gap'' 
would close at the merger point, and then, the flux cannot pass to the opposite side. 
This implies that the topological fluctuation is not suppressed at the merger point.%
\footnote{
It should be noticed that the stringy effects would become important
very near the merger point due to the large curvature.
However our arguments in this section relies only on the topological properties of the geometries  and our results would be valid as far as the topologies are well defined.
An important question about the merger point is whether the singularity is resolved by stringy and/or quantum gravity effect.
Our result shows that the D1-branes become light near the merger point and 
it suggests that the D1-branes may contribute to the singularity resolution.
}

Recall that the merger point will correspond to the Gross-Witten-Wadia type transition point in the gauge theory  \cite{Gross:1980he, Wadia:1980cp, Wadia:2012fr}, where the topology of the eigenvalue distribution of the Polyakov loop operator changes.
It indicates that the topology of the eigenvalue distribution 
of the Polyakov loop is crucial for the topological fluctuation in QCD.
It sounds reasonable since both the Polyakov loop and instantons 
are related to the configurations of the gauge fields.

On the other hand, the translation symmetry along $t'$-circle, which is broken in both the localized D3-brane and the non-uniformly smeared D3-brane geometry, is not critical  for the topological fluctuation.
This translation symmetry correspond to the $Z_N$ symmetry, which characterizes the confinement \cite{Mandal:2011ws}, and we predict that this symmetry is not directly connected to the large fluctuation of the topological charge.

\section*{Acknowledgements}
The authors would like to thank Robert Pisarski and Edward Shuryak for stimulating discussions and comments.
The authors also would like to thank Tatsuo Azeyanagi and Shotaro Shiba for collaboration in the initial stages of this project.
T. M. would like to thank Yukawa Institute for hospitality where part of the work was done.
The work of M.~H. is supported in part by the Grant-in-Aid of the Japanese Ministry of Education, 
Sciences and Technology,  Sports and Culture (MEXT) for Scientic Research (No.  25287046).
The work of Y.~M. is supported in part by European Union's Seventh Framework 
Programme under grant agreements (FP7-REGPOT-2012-2013-1) no 316165, 
the EU program ``Thales'' MIS 375734 and was also cofinanced by 
the European Union (European Social Fund, ESF) and Greek national funds through
the Operational Program ``Education and Lifelong Learning'' of 
the National Strategic Reference Framework (NSRF) under 
``Funding of proposals that have received a positive evaluation 
in the 3rd and 4th Call of ERC Grant Schemes.''
The work of T.~M. is supported in part by Grant-in-Aid for Scientific Research (No. 15K17643) from JSPS.

\appendix

\section{Stable configuration of D1-brane in the localized D3-brane geometry}
\label{sec-instanton-stability}
In section \ref{sec-large-instanton}, we calculated the DBI action for 
the stable configuration of the D1-brane, 
in which the D1-brane is stretched between the D3-branes and their mirror images at $u=0$. 
Here, we argue that this configuration is the only stable configuration 
of the D1-brane which wraps on $t'$- and $x_4$-directions.

We consider the D1-brane which is embedded in the three-dimensional space of $(u,t',x_4)$. 
To investigate the D1-brane in this space, we express the background metric (\ref{metric-unified}) as 
\begin{equation}
 ds^2 = 
 \alpha' \left[
 H^{-1/2} \left( \sum_{i=1}^{3} dx_i^2 
 + f_4 dx_4^2  \right)
 + H^{1/2} f_r \left(d r^2 + r^2 d \theta^2 +	r^2 \sin^2\theta d \Omega_4^2\right) 
 \right] \ ,  
\end{equation}
and choose the coordinate $\theta$ so that the  three-dimensional space of $(u,t',x_4)$ is parameterized by $(r,\theta,x_4)$, and $\theta=0$ (and $\theta=\pi$) corresponds to $u=0$.
Then the D1-brane lies on $(r,\theta,x_4)$ space and the induced metric on it is given by 
\begin{equation}
 ds^2 = 
 \alpha' \left[
 H^{-1/2} f_4 dx_4^2 
 + H^{1/2} f_r \left(1+r^2 \theta^{\prime 2}\right)  dr^2 
 \right] \ ,  
\end{equation}
where $\theta$ is a function of $r$ and $\theta' = d\theta / dr$. 
The DBI action can be expressed as 
\begin{equation}
 S_\text{D1} = 2 \frac{N \beta}{ \lambda_5 } 
 \int_{r_s}^{\beta'/2} d r \int_{0}^{L_4} dx_4 \sqrt{f_4f_r(1+r^2 \theta^{\prime 2})} \ . 
\end{equation}
Since $\theta$ dependence of $f_4$ and $f_r$ are negligible if $r_0 \ll \beta'$ as we argued in section \ref{sec-large-instanton}, we can solve the equation of motion for $\theta$ as 
\begin{equation}
 \theta' = \frac{c_1}{r\sqrt{r^2 f_4 f_r -c_1^2}} \ ,
 \label{eom-theta}
\end{equation}
where $c_1$ is a constant.

\begin{figure}
\begin{center}
\includegraphics[scale=1.0]{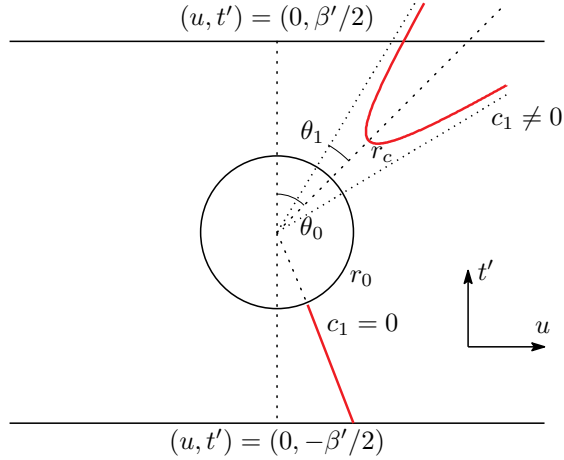}
\caption{Schematic plots of the D1-brane in ($u,t'$) space.
The red lines describe the D1-brane which obeys the equation of motion (\ref{eom-theta}). 
}
\label{fig-sol}
\end{center}
\end{figure}

If $c_1=0$, $\theta$ becomes a constant, and the solution describes the D1-brane which is orthogonal to the ``horizon'' at $r=r_0$ and extends straightly to outside with a fixed angle $\theta_0$. (See Fig.~\ref{fig-sol}.)
If we choose $\theta=0$ (and $\pi$) so that the D1-brane lies along $u=0$, we obtain the stable solution which we investigated in section \ref{sec-large-instanton}. 

If $c_1 \neq 0$, the brane is curved in the $(r,\theta)$-plane.
For small $c_1$ ($\ll r_0$), the solution behaves around $r=r_0$ as 
\begin{equation}
 \theta = \theta_0 +  \frac{\sqrt{c_1}}{2^{1/4}r_0} \left(\sqrt{r - r_c} + \cdots \right)  , \qquad
 r_c =  r_0 + \frac{1}{2^{3/2}} c_1 
+ \mathcal O(c_1^2)
 \label{sol-dbi-near}
\end{equation}
This solution describes the D1-brane which does not reach to $r=r_0$, 
but turns at $r=r_c$  and goes back to the outside of the near region. (See Fig.~\ref{fig-sol}.)
Since (\ref{eom-theta}) indicates $\theta' \to 0$ as $r \to \infty$, the D1-brane asymptotically extends to angles $\theta_0 \pm \theta_1$ as $r \to \infty$ where the asymptotic value $ \theta_1$ is fixed by $c_1$.
The solution (\ref{sol-dbi-near}) indicates that $\theta_1$ will decrease as $r_c$ decreases and achieves 
$\theta_1=0$ at $r_c=r_0$ ($c_1=0$).
Indeed we can confirm that $\theta_1$ approaches to the maximum $\pi/2$ as $r_c\to\infty$ by solving (\ref{eom-theta}) explicitly. 

So far we have not considered the periodicity of $t'$-cycle, and now we impose it to the solutions.
We demand that the solutions are smoothly connected at $t'=\pm \beta'/2$ which are the middle points between the D3-brane and its mirrors. 
This leads the following boundary conditions;
\begin{align}
 U(t'= \pm \beta'/2) = U_0 \ , \qquad
 \left.\frac{dU(t')}{dt'}\right|_{t'=  \pm\beta'/2} =0\ , 
 \label{BC}
\end{align}
where we have taken the coordinates $(u,t')$ and $U(t')$ is the profile of the D1-brane in these coordinates.
Then we immediately notice that the possible solutions are $c_1=0$ with $\theta=0$ and $c_1 \neq 0$ with $\theta_0=\pi/2$ and $\theta_1 = \pi/2 $ ($r_c = \infty$) only.
These are the constant $u$ solutions at $u=0$ and $u= \infty$ respectively.
Thus $u=0$ is the only stable solution of the D1-brane in the localized D3-brane geometry.

However, the higher order corrections of $r_0$ might be relevant in the intermediate region between the asymptotic region and near region. 
Although the D1-brane is approximated by straight configuration in the asymptotic region 
in the above analysis, 
the higher order correction may bend the D1-brane and it might allow other solutions.
In order to be a solution which satisfies the boundary condition \eqref{BC} with  $U_0\neq 0$, 
it must go toward the D3-brane from $t'=\beta'/2$. 
Let us see whether it happens.
The DBI action of the D1-brane in the asymptotic region is given by \eqref{dbi-asympt}. 
By assuming $U'\ll 1$, the equation of motion becomes 
\begin{equation}
 \frac{d^2 U(t')}{dt^{\prime 2}} 
 = 
 \left.\frac{3}{4} \frac{\partial\Phi}{\partial u} \right|_{u=U(t')}
\end{equation}
We solve this equation around $t'=\beta'/2$ with the boundary condition (\ref{BC}) and obtain
\begin{equation}
 U(t') 
 \sim 
 U_0 + \left.\frac{3}{8} \frac{\partial\Phi}{\partial u} 
 \right|_{\substack{u=U_0 \\ t'=\beta/2}} t^{\prime 2}
 + \cdots \ . 
\end{equation}
Since $\partial_u \Phi > 0$ for $u\neq 0$ from (\ref{metric-LD3-large-u}), 
the D1-brane goes away from the D3-branes for $U_0\neq 0$. 
Therefore the higher order corrections of $r_0$ does not change the result.
Hence the D1-brane located at $u=0$ is the only stable configuration even if we take into account the corrections of $r_0$.

It would be worth comparing with the case of D3-D7 system \cite{Mateos:2007vn} 
in which the D7-brane has non-trivial stable configurations. 
In this case, the D7-brane is not straight outside the horizon 
and approaches to $U=\text{const}$. 
This is because the D7-brane wraps on the $S^3$ and 
hence tends to stay in the region with small radius due to the tension of these directions. 
On the other hand, the D1-brane does not wrap no cycle other than 
$x_4$ and $t'$ and hence extends straightly in the near region.


\end{document}